\def\NN{\nonumber}

\def\r{\right}
\def\l{\left}
\def\R{\rangle}
\def\L{\langle}

\def\kv{{\bf k}}
\def\om{\omega}
\def\eps{\varepsilon}
\def\Gc{{\cal G}}

\def\Om{\Omega}
\def\dg{\dagger}
\def\Vh{{\hat V}}
\def\Gh{{\hat G}}
\def\e{{\rm e}}
\def\ZZ{{\Bbb Z}}
\def\s{\sigma}
\def\up{\uparrow}
\def\dn{\downarrow}
\def\sumind#1#2{{\scriptstyle #1\atop\scriptstyle #2}}

\documentstyle[prl,aps,twocolumn,amsfonts]{revtex}
\input epsf.tex
\begin{document}
\twocolumn[
\hsize\textwidth\columnwidth\hsize\csname@twocolumnfalse\endcsname

\title{The spectral weight of the Hubbard model through
cluster perturbation theory}

\author{D. S\'en\'echal, D. Perez and M. Pioro-Ladri\`ere}

\address{Centre de Recherche en Physique du Solide et D\'epartement de
Physique,}
\address{Universit\'e de Sherbrooke, Sherbrooke, Qu\'ebec,
Canada J1K 2R1.}
\date{August 1999}

\maketitle
\begin{abstract}
We calculate the spectral weight of the one- and two-dimensional Hubbard
models, by performing exact diagonalizations of finite clusters and treating
inter-cluster hopping with perturbation theory. Even with relatively modest
clusters (e.g. 12 sites), the spectra thus obtained give an accurate
description of the exact results. Thus, spin-charge separation (i.e. an extended
spectral weight bounded by singularities) is clearly recognized in the
one-dimensional Hubbard model, and so is extended spectral weight in the
two-dimensional  Hubbard model.
\end{abstract}

]

One of the central issues in the theory of strongly correlated electrons is the
existence or not of well-defined quasiparticles. This question is best
addressed by studying the spectral weight (SW) $A(\kv,\om)$, i.e., the
probability distribution  for the energy $\hbar\om$ of an electron of
wavevector $\kv$ added to, or removed from the system. In a Fermi liquid, the
SW is dominated by a single quasiparticle peak centered at
$\om=\eps(\kv)$, whose width decreases as $\eps(\kv)$ approaches the Fermi
energy. In a Luttinger liquid, the SW  is distributed between two
singularities associated respectively to spin and charge
excitations (spinons and holons)\cite{Voit94}. The hole (i.e.,
electron-removal) part of the SW can be measured by angle-resolved
photoemission spectroscopy (ARPES), a technique which has improved steadily in
recent  years\cite{Kim96,Kim98,Ronning98}. On the theoretical side, the SW is
obtained from the one-particle Green function:
\begin{equation}
\label{SWdef}
A(\kv,\om)=\lim_{\eta \to 0^+}
-2\>{\rm Im}\>\Gc(\kv,\om +i\eta),
\end{equation}
and the latter may be approximately evaluated by various analytical and
numerical methods.

In this letter we explain a new method for calculating $A(\kv,\om)$ in
Hubbard-type models, based on a combination of exact diagonalizations (ED) of
finite clusters with strong-coupling perturbation
theory\cite{Pairault98,Pairault99}, and apply it to the one- and
two-dimensional Hubbard models. Exact diagonalizations based on the
Lanczos algorithm are commonly used to evaluate
$A(\kv,\om)$\cite{Dagotto94,Leung92,Otha92,Favand97}. Unfortunately,
computer memory requirements grow exponentially with system size and
restrict the analysis to small clusters (e.g. 16 sites for the Hubbard
model). The SW thus obtained is the sum of a relatively small number of
poles and its extended character in the thermodynamic limit is
difficult to assess. The SW may also be evaluated by quantum Monte
Carlo (QMC)\cite{Bulut94,Preuss94,Preuss95,Zacher98}: larger systems
may thus be studied (e.g. 64 sites) but the maximum entropy method
(MEM) used for approximate analytic continuation tends to produce
smooth SW and may miss weak features; moreover, computation time
increases as the temperature is lowered. The new method we propose
consists in (i) dividing the lattice into identical $N$-site clusters;
(ii) evaluating -- by ED -- the one-particle Green function
$G_{a,b}(z)$ within a cluster ($a,b$ are lattice sites and $z$ a complex
frequency) with open boundary conditions; (iii) treating the inter-cluster
hopping $t_0$ in perturbation theory and recovering the Green function
$\Gc(\kv,\om)$. Thus, short-distance effects are treated exactly, while
long-distance propagation is treated at the RPA level.

\begin{figure}
\epsfxsize 5truecm\centerline{\epsfbox{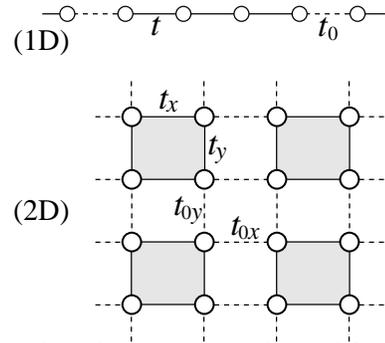}} 
\caption{Dividing the lattice into identical four-site clusters for the 1D and
2D Hubbard models. The in-cluster hopping amplitude is $t$ and the
inter-cluster hopping is $t_0$.}
\label{clusterFIG}
\end{figure}

Step (i) is illustrated on Fig.~\ref{clusterFIG}. We denote by $t$ and $t_0$
the hopping amplitudes within and between clusters, respectively ($t$ and
$t_0$ may be different a priori, but will be identical in practice).
Clusters of up to $N=12$ sites have been treated, with various aspect ratios in
the 2D case. Open (free) boundary conditions
must be used.
Step (ii) proceeds according to
the usual Lanczos method\cite{Dagotto94}. The cluster Green function
$G_{a,b}(z)$ is defined as
\begin{equation}
G_{a,b}(z) = \L\Om|c_a{1\over z-H}c^\dg_b|\Om\R + 
\L\Om|c^\dg_b{1\over z+H} c_a|\Om\R
\end{equation}
where $|\Om\R$ is the ground state obtained by ED, $c_a$ is the electron
destruction operator at site $a$ (we drop the spin index) and $H$ is the
Hamiltonian (including chemical potential). The two terms in $G_{a,b}$
correspond respectively to electron ($G_{a,b}^{\rm e}$) and hole
($G_{a,b}^{\rm h}$) propagation, and are calculated separately. In the
subspace containing one additional electron (with respect to the ground
state), an accurate tridiagonal representation of $H$ is obtained (typically
of dimension ranging from 50 to 250). Efficient routines for inverting
tridiagonal matrices are used to evaluate
$G_{a,b}^{\rm e}(z)$ at any desired complex value, and likewise for
$G_{a,b}^{\rm h}(z)$. In usual Lanczos calculations, this inversion takes the
form of a continued-fraction representation.

Step (iii) demands more explanations. Let $c_{m,a}$ be the electron destruction
operator on site $a$ of cluster $m$ ($a=1,\dots,N$). The full system is treated
as a superlattice of clusters, each cluster being made of $N$ ordinary lattice
sites; we will work in one dimension for simplicity, but a suitable
generalization to higher-dimensional lattices is readily obtained.  The complete
Hamiltonian of the system may be written as
$H=H_0+V$:
\begin{equation}
H_0 = \sum_{m\in\ZZ} H^0_m \quad,\quad
V = \sum_\sumind{m,n}{a,b} V^{m,n}_{a,b}
c_{m,a}^\dg c_{n,b}
\end{equation}
where $H^0_m$ is, say, the Hubbard Hamiltonian of the $m^{\rm th}$ cluster
\begin{eqnarray}
H^0_m &=& -t\sum_{\L a,b\R}\sum_{\s} \l(
c^\dg_{m,a,\s}c_{m,b,\s}+{\rm H.c.} \r)
\NN\\&&\qquad
+ U\sum_a n_{m,a,\up}n_{m,a,\dn}
\end{eqnarray}
and $V$ is the nearest-neighbor hopping between adjacent clusters:
\begin{equation}\label{Vmn}
V^{m,n}_{a,b} = -t_0(\delta_{m,n-1}\delta_{a,N}\delta_{b,1} +
\delta_{m,n+1}\delta_{a,1}\delta_{b,N})
\end{equation}
Of interest is the electron Green function $\Gc_{a,b}(Q,z)$, where $Q$ is a
superlattice wavevector, and $a,b$ are site indices within a cluster.
The perturbation $V$ being a one-body operator, it may be treated in the
formalism of Refs~\cite{Pairault98,Pairault99}, wherein a systematic
perturbation expansion was constructed for such terms. The lowest-order
contribution to this expansion has the RPA form:
\begin{equation}\label{RPA1}
\Gc_{a,b}(Q,z) = \l({\Gh(z)\over 1-\Vh(Q)\Gh(z)}\r)_{a,b}
\end{equation}
where $\Gh(z)$ is the generalization of the ``atomic'' Green function, now a
$N\times N$ matrix in the space of site indices. Likewise, $\Vh(Q)$ is the
reciprocal superlattice representation of the hopping (\ref{Vmn}):
\begin{equation}\label{VQ}
V_{a,b}(Q) =
-t_0\l(\e^{iQ}\delta_{a,N}\delta_{b,1}+\e^{-iQ}\delta_{a,1}\delta_{b,N}\r)
\end{equation}
Relation (\ref{RPA1}) may be regarded as a cluster generalization of
the Hubbard-I approximation\cite{Hubbard63a}.

The Green function $\Gc_{a,b}(Q)$ of Eq.~(\ref{RPA1}) is in a mixed
representation: real space within a cluster and Fourier space between clusters.
A true Fourier representation in terms of the original reciprocal lattice is
preferred. Since the cluster decomposition breaks translation invariance, $\Gc$
will depend on two continuous momenta $k$ and $k'$, identical modulo a reciprocal
superlattice vector:
\begin{eqnarray}
\Gc (k,k';z) &=& {1\over N}\sum_{s=0}^{N-1}
\delta\l(k-k'+{2\pi s\over N} \r)
\NN\\ &&\quad\times
\sum_{a,b=1}^N\e^{-ik(a-b)}
\e^{2\pi i sb/N}\Gc_{a,b}(Nk,z)
\end{eqnarray}
If we set $t=t_0$, the $s=0$ component ($k=k'$) is the RPA approximation to the
full Green function:
\begin{equation}\label{RPA2}
\Gc_{\rm RPA} (k,z) = {1\over N} \sum_{a,b}\e^{-ik(a-b)}\Gc_{a,b}(Nk,z)
\end{equation}
Eqs~(\ref{RPA1},\ref{RPA2}) are then used to calculate the SW. 

The approximation (\ref{RPA1}) turns out to be exact in the absence of
interactions ($U=0$). In that case, Wick's theorem applies and the RPA form is
the exact resummation of the perturbation series, $\Vh(Q)$ being the exact
self-energy. If we set $t=t_0$, Eq.~(\ref{RPA2}) then yields the exact
Green function for an infinite system at arbitrary wavevector, from the exact
Green function $G$ of an finite, open cluster. When $U\ne0$,
Expression~(\ref{RPA2}) is no longer exact, but strong interactions tend to
cause short-range correlations that are incorporated with good accuracy in
modest-size clusters. Thus short-distance effects are well served by the ED
within a cluster, and long-distance effects by the RPA approximation, making
our method adequate at intermediate coupling.

\begin{figure}
\epsfxsize 6truecm\centerline{\epsfbox{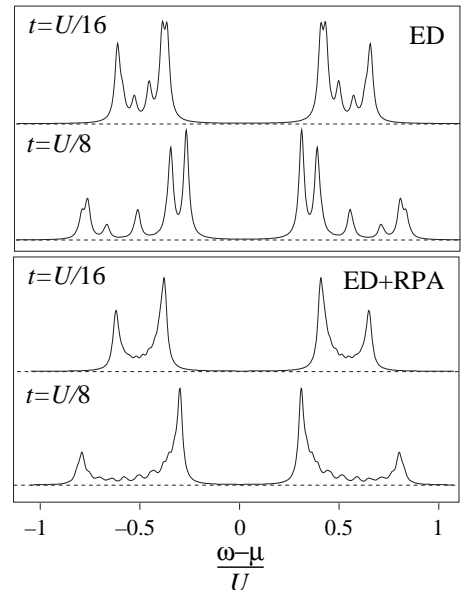}} 
\caption{
Above: $A(\pi/2,\om)$ for the 1D Hubbard model at half-filling on a 12-site
ring with ordinary ED. Below: the same, but with applying
RPA with 12-site clusters. The extended character of the SW is manifest. Note: the parameter $\eta$ of Eq.~(\protect\ref{SWdef})
has been set to 0.03 in order to give delta peaks a finite width.}
\label{ED-RPAFIG}
\end{figure}
We have applied the method just described to the 1D Hubbard model.
Fig.~\ref{ED-RPAFIG} compares the result of an ordinary ED on a 12-site ring
with the present approach: Whereas the extended character of $A(k,\om)$ --
here a signature of spin-charge separation -- is not clear form the ED data
alone, it is clearly revealed by the RPA spectra. In fact, the two
branches of the SW can already be seen with a two-site cluster (not shown),
but more and more poles appear in between when the cluster size is increased:
the actual separation of spin and charge excitations needs a fair cluster size
to occur, and propagation between clusters at the RPA level requires the holon
and spinon to recombine.

\begin{figure}
\epsfysize 8.8truecm\centerline{\epsfbox{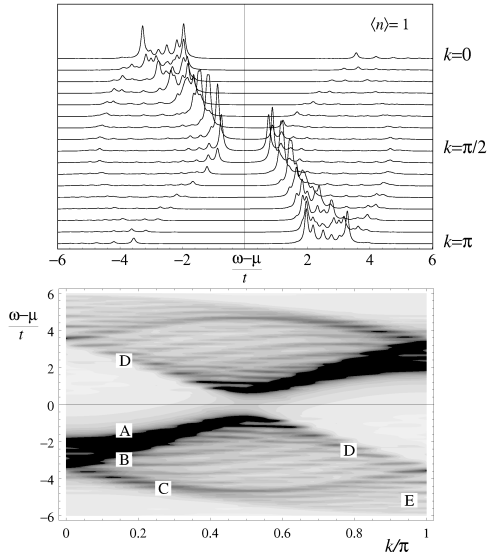}} 
\caption{
Spectral weight of the 1D Hubbard model at half-filling, for $U=4t$,
calculated from Eqs.~(\protect\ref{RPA1}) and (\protect\ref{RPA2}) with $N=12$.
Below, density plot of the same data.
}
\label{1DAFIG}
\end{figure}
Fig.~\ref{1DAFIG} illustrates the SW of the 1D Hubbard model at
half-filling with $U/t=4$. Most noteworthy are: (i) The extended character
of the SW, with six branches having a clear dispersion, even though
most of the weight lies near the ``quasiparticle band'' following
approximately the $-2t\cos k$ free-particle dispersion; (ii) the gap
opening at $k=\pi/2$; (iii) the spinon (A) and holon (B) branches,
characteristic of a Luttinger liquid with a charge gap (Branch D is the mirror
of the holon branch with opposite frequency)\cite{Voit98}; (iv) the weak,
higher-frequency band (C), absent from low-energy Luttinger liquid
predictions; (v) the high-frequency tail (E) near the zone boundary.
Band C, as well as Bands B and D together, disperse with period
$\pi$ instead of $2\pi$, a signature of local AF correlations. A comparison
with Fig.~1c of Ref.~\cite{Favand97} -- which illustrates the SW in the
$U\to\infty$ limit -- is revealing of the changes brought about by a finite
value of $U$: in the $U\to\infty$ limit, just the hole part of the SW is
present, but the same branches can be found, however with comparable
relative intensity: Branches D and E are the mirror images of branches B and A,
respectively. The finite value of $U$ weakens considerably the intensity of
branches C, D, and E.  It is also interesting to compare Fig.~\ref{1DAFIG}
with Fig.~2 of Ref.~\cite{Preuss94} and Fig.~3 of Ref.~\cite{Pairault98},
where the same parameters were used. In particular, it is clear that the
Maximum Entropy Method of Ref.~\cite{Preuss94} lumps the spinon and holon
peaks near $k=0$ into one broad peak.

\begin{figure}
\epsfysize 8.8truecm\centerline{\epsfbox{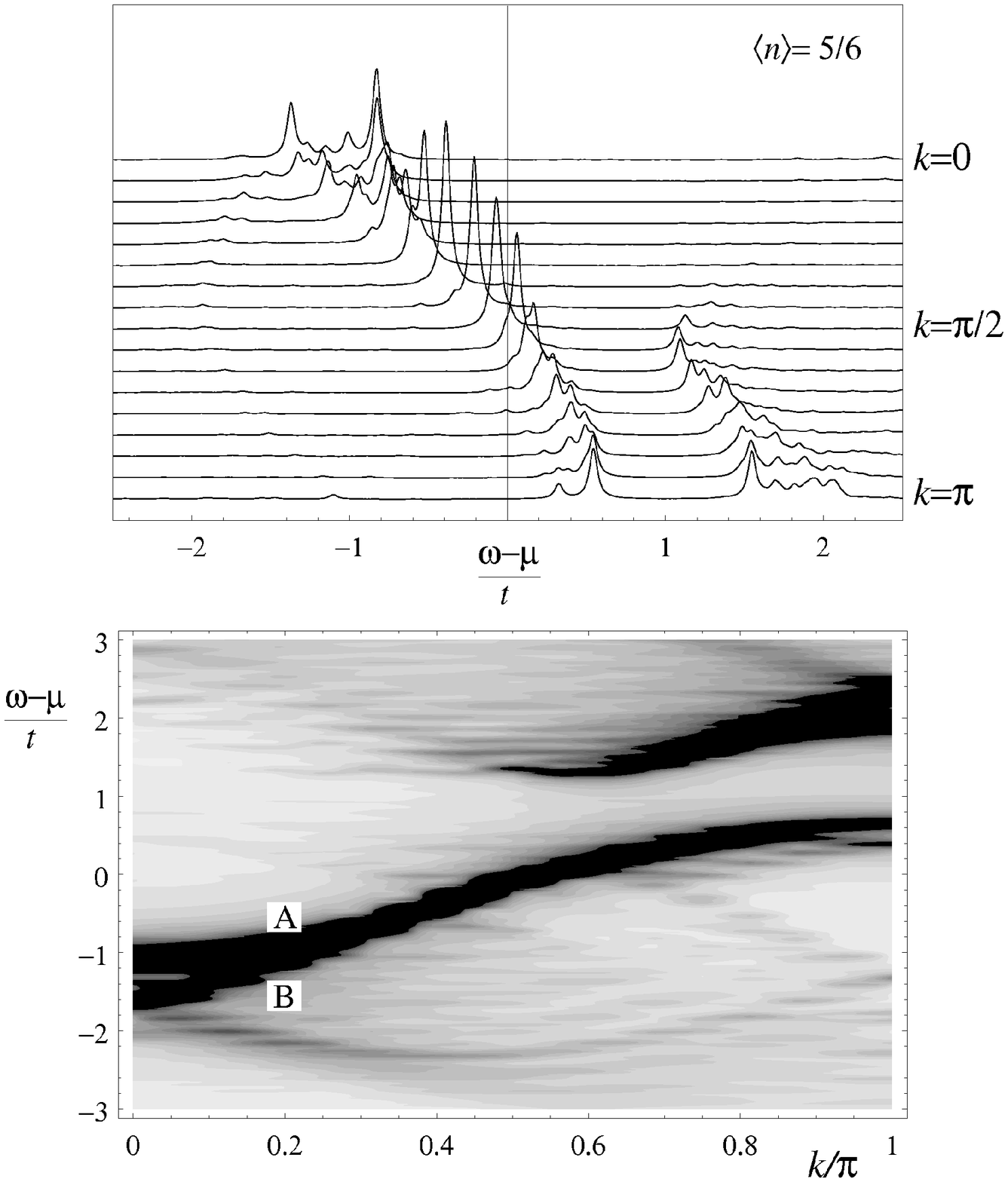}} 
\caption{
Spectral weight of the 1D Hubbard model at $\L n\R={5\over6}$ and $U=4t$,
calculated from Eqs.~(\protect\ref{RPA1}) and (\protect\ref{RPA2}) with $N=12$.
Below: density plot of the same data.
}
\label{1DBFIG}
\end{figure}
Fig.~\ref{1DBFIG} shows the SW of the 1D Hubbard model for the same value of
$U/t$, but away from half-filling, at $\L n\R={5\over6}$. The chemical
potential $\mu=0.64$ was inferred from the integrated density of states. The
Fermi level crosses the main band, causing metallic behavior, but again the SW
is clearly extended, with a clear weakening of the upper Hubbard band: there
is a significant transfer of SW from high to low frequencies\cite{Meinders93}.
Again, the spinon (A) and holon (B) branches are clearly identified, this time
in a gapless Luttinger liquid. This may be compared with Fig.~3 of
Ref.~\cite{Zacher98}, which corresponds to the same $U/t$ ratio, but with $\L
n\R={3\over 4}$. 

\begin{figure}
\epsfysize 8.8truecm\centerline{\epsfbox{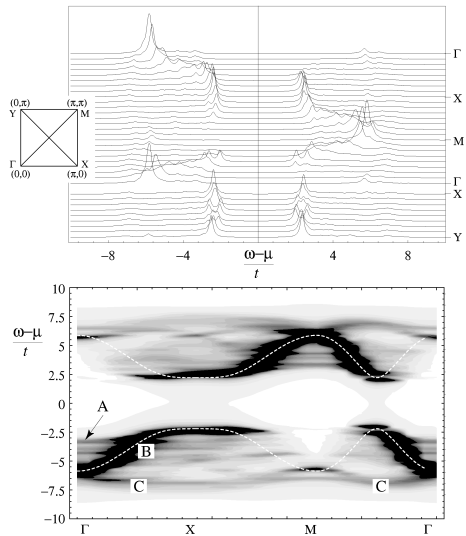}} 
\caption{
Spectral weight of the 2D Hubbard model at half-filling, for $U=8t$,
calculated from Eqs.~(\protect\ref{RPA1},\protect\ref{RPA2}) with a $3\times4$
cluster. Inset: The wavevector scans. Below: density plot of the same data
(except for the scan from X to Y) to be compared directly with Fig.~1 of
Ref.~\protect\cite{Preuss95}. The dashed curve is the best SDW dispersion,
with $t=1.36$ and gap $\Delta=2.21$.
}
\label{2DAFIG}
\end{figure}

We have also applied our method to the 2D Hubbard model, with various
cluster shapes ($2\times6$, $3\times4$, $4\times3$ and $6\times2$). The
spectra obtained from these different shapes are very similar, and
this reinforces our confidence in them, even though the linear dimensions
of the clusters are modest. Fig.~\ref{2DAFIG} illustrates the SW of the 2D
Hubbard model at half-filling for $U=8t$, with a $3\times4$ cluster. This is
to be compared with Fig.~1 of Ref.~\cite{Preuss95} and Fig.~6 of
Ref.~\cite{Pairault99}. In contrast to the 1D case, the SW is much more
concentrated around one peak, but its extended character is still undeniable.
Indeed, one is tempted to draw an analogy with 1D spinon and holon branches:
the momentum scan $\rm\Gamma-X-M$ shares features with the
$[0,\pi]$ scan in the 1D case, except that the ``spinon'' (A) is much weaker
than the ``holon'' (B). The same can be said of the diagonal scan $\rm\Gamma-M$
(from right to left on Fig.~\ref{2DAFIG}). Likewise, a high-frequency
band (C) is visible. Most obvious is the gap opening at the Fermi surface,
constant along the XY line, a feature that would certainly be modified by
including a diagonal hopping $t'$, and which demonstrates anyway that
nearest-neighbor hopping alone cannot account for the ARPES data of insulating
cuprates\cite{Ronning98} (this was already known for the $t-J$ model). Note,
that whereas Refs.~\cite{Preuss95} and\cite{Pairault99} both resolve two peaks
near Point M, the present approach suggests an extended SW at that point.
The expected antiferromagnetic order of the half-filled 2D Hubbard
model is not seen here, because of finite cluster size. This order would imply
a folding of the Brillouin zone, with a corresponding symmetry of the SW
following the SDW fit illustrated on Fig.~\ref{2DAFIG}.

Some general remarks are in order. Formulas (\ref{RPA1},\ref{RPA2}) are
but the first order result of a systematic $t_0$ expansion
(See Ref.~\cite{Pairault99} for details). It is difficult to assess the
convergence of this perturbative expansion, since it depends certainly on the
ratio $t/U$ and on cluster size $N$.  We expect nonetheless the method to give
better results at strong coupling, where short-range effects dominate and are
thus well accounted for by modest clusters. Indeed, the effect of
antiferromagnetic correlations are already seen with two-site clusters. Going
to order $t_0^2$ in strong-coupling perturbation is a way of improving the
results presented here, but appears quite difficult in practice, because of
the need to compute numerically exact two-particle Green functions on a
cluster.
The spectra presented here are all normalized, up to 1 or 2\%. The general form
(\ref{RPA1}) guarantees that the continued fraction form of the SW
will have the correct first coefficient, thus ensuring its normalization.

In summary, we have shown how strong-coupling perturbation theory can be used
to incorporate long-distance effects into ED data which already contain
short-distance effects exactly. This method allows for a clear recognition of
spin-charge separation in the 1D Hubbard model, and of extended SW in the 2D Hubbard model. Further applications of this method (NNN
hopping, three-band Hubbard model, etc.) are under way.

%
We thank S.~Pairault and A.-M. S. Tremblay for numerous enlightening
discussions. This work was partially supported by NSERC (Canada) and FCAR
(Qu\'ebec).


\end{document}